\documentclass[sigconf]{acmart} 

\AtBeginDocument{%
  \providecommand\BibTeX{{%
    \normalfont B\kern-0.5em{\scshape i\kern-0.25em b}\kern-0.8em\TeX}}}

\copyrightyear{2024}
\acmYear{2024}
\setcopyright{rightsretained}
\acmConference[CHI EA '24]{Extended Abstracts of the CHI Conference on Human Factors in Computing Systems}{May 11--16, 2024}{Honolulu, HI, USA}
\acmBooktitle{Extended Abstracts of the CHI Conference on Human Factors in Computing Systems (CHI EA '24), May 11--16, 2024, Honolulu, HI, USA}
\acmDOI{10.1145/3613905.3650783}
\acmISBN{979-8-4007-0331-7/24/05}

\acmSubmissionID{5510}

\author{Yihan Hou}
\email{yhou073@connect.hkust-gz.edu.cn}
\affiliation{%
  \institution{The Hong Kong University of Science and Technology (Guangzhou)}
  \city{Guangzhou}
  \country{China}
}

\author{Hao Cui}
\email{hcui058@connect.hkust-gz.edu.cn}
\affiliation{%
  \institution{The Hong Kong University of Science and Technology (Guangzhou)}
  \city{Guangzhou}
  \country{China}
}

\author{Rongrong Chen}
\email{rainerrchen@uic.edu.cn}
\affiliation{%
  \institution{Guangdong Provincial Key Laboratory IRADS, BNU-HKBU United International College}
  \city{Zhuhai}
  \country{China}
}

\author{Wei Zeng}
\email{weizeng@hkust-gz.edu.cn}
\authornote{Wei Zeng is the corresponding author}
\affiliation{%
  \institution{The Hong Kong University of Science and Technology (Guangzhou)}
  \city{Guangzhou}
  \country{China}
}
\affiliation{%
  \institution{The Hong Kong University of Science and Technology}
  \city{Hong Kong SAR}
  \country{China}
}
\begin{document}

\title{Understanding the Impact of Referent Design on Scale Perception in Immersive Data Visualization}


\newcommand{\re}[1]{\textcolor{myblue}{#1}}
\definecolor{myblue}{RGB}{0, 0, 0}


\begin{abstract}
    \label{sec:abstract}
    Referents are often used to enhance scale perception in immersive visualizations.
    Common referent designs include the considerations of referent layout (\emph{side-by-side} vs. \emph{in-situ}) and referent size (\emph{small} vs. \emph{medium} vs. \emph{large}).
    This paper introduces a controlled user study to assess how different referent designs affect the efficiency and accuracy of scale perception across different data scales, on the performance of the size-matching task in the virtual environment.
    Our results reveal that in-situ layouts significantly enhance accuracy and confidence across various data scales, particularly with large referents. Linear regression analyses further confirm that in-situ layouts exhibit greater resilience to changes in data scale. For tasks requiring efficiency, medium-sized referents emerge as the preferred choice.
    Based on these findings, we offer design guidelines for selecting referent layouts and sizes in immersive visualizations.
\end{abstract}






\maketitle

\section{Introduction}
\re{Advancements in virtual reality (VR) have led to immersive data experiences that enable the concrete representation of abstract data, aiming for straightforward interpretation, especially in immersive data storytelling.
In this context, the reliance on semantic marks leads to the absence of axes or magnitudes, posing challenges in accurately conveying data scales~\cite{yang2023understanding, lee2020data}.
This issue is exacerbated by VR's inherent size distortion, where objects tend to appear smaller~\cite{thompson2004does, kenyon2007size, kelly2017perceived}, especially at longer ranges~\cite{hornsey2020size}.}
Many studies have been devoted to assessing size distortion and its influencing factors in VR~\cite{park2021judgments, wijayanto2023comparing}.
The anchor effect, which involves the incorporation of referents, has been recognized as a promising method to provide precise size perception~\cite{ogawa2018object,ogawa2019virtual}.
Similarly, immersive visualization can also be enhanced by incorporating referents to alleviate size distortion.

While the space for referent design has been explored~\cite{chevalier2013using}, there is an absence of empirical evidence regarding the effectiveness of referent design in immersive visualizations.
To address the gap, we conduct a within-subjects comparative analysis to investigate the influence of referent design on scale perception in immersive visualization.
Specifically, we examine the effects of referent size (\textit{small} vs. \textit{medium} vs. \textit{large} with reference to the human body) and the layout configuration (\emph{in-situ} vs. \textit{side by side}) on scale matching tasks across different data scales.
The evaluation primarily centers on the impact of these parameters on task completion efficiency, accuracy, and self-reported confidence.
The results reveal that \emph{in-situ} layout significantly outperforms \emph{side-by-side} layout, demonstrating higher accuracy and task completion confidence across all referent sizes.
Notably, with the increase in data scale, \emph{in-situ} layout exhibits greater resilience in scale perception compared to \emph{side-by-side} layout.
Moreover, the referent of \emph{medium} size generates faster completion time than those of \emph{small} and \emph{large} sizes.
Building upon these findings, we suggest recommendations for referent design in immersive visualizations.
\section{Related Works}

\begin{figure*}[htb]
    \centering
    \includegraphics[width=0.95\linewidth]{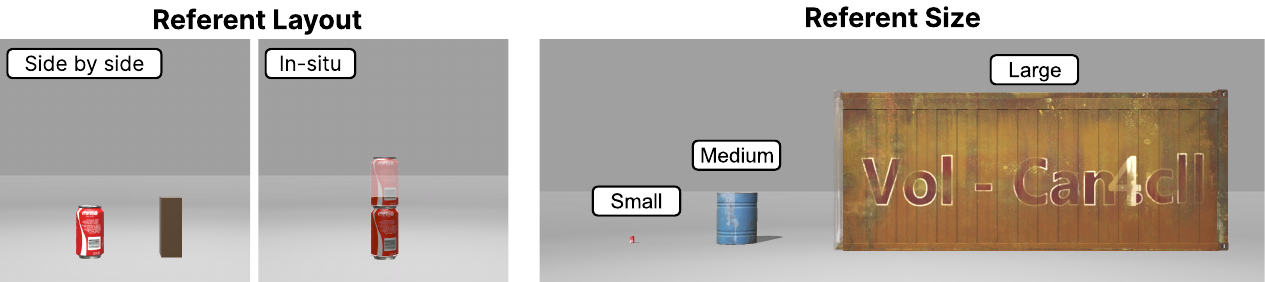}
    \caption{Design parameters of the referent design that are considered in the study. The Left shows two variations of the referent layout (in-situ vs. side-by-side), and the right shows three variations of the referent size (small vs. medium vs. large).}
    \Description{The figure shows two design parameters of the referent design that are considered in the study, namely, the referent layout and the referent size. The left side of the figure shows two variations of the referent layout, in-situ and side-by-side. The right side of the figure shows three variations of the referent size: small, medium, and large. Details are provided in the text.}
    \label{fig:design_parameter}
\end{figure*}

\subsection{Scale Perception in Virtual Reality}
Scale perception plays a critical role in spatial understanding within virtual environments, yet it is subject to size distortions.
Studies have found that objects tend to appear smaller, particularly at increased distances~\cite{thompson2004does, kenyon2007size, kelly2017perceived,hornsey2020size}.
A broad spectrum of influencing factors has been identified, including eye height~\cite{mine2020relationship, arora2021thinking}, scene complexity~\cite{luo2007effects, park2021judgments}, visual realism~\cite{wijayanto2023comparing}, optic flow~\cite{jones2013peripheral}, sensory-perceptual channels~\cite{de2021empirically}, \re{physics simulations}~\cite{pouke2020plausibility, pouke2021plausibility}, and viewing angle~\cite{weber2020body}.
Interestingly, despite the presence of distortions, individuals exhibit consistent size perception of familiar objects within virtual environments~\cite{distler2000velocity, nguyen2008effects, linkenauger2013welcome, rzepka2023familiar}.
This suggests that these familiar objects can serve as benchmarks for estimating the size of unfamiliar ones.
\re{Beyond such visual anchors, body-based scaling emerges as a pivotal factor in size perception, with our body acting as a fundamental reference in assessing size~\cite{van2011being}.
    Notably, embodying an avatar with altered proportions can reshape our perception of the world around us~\cite{ogawa2018object, ogawa2019virtual, linkenauger2013welcome, mine2020relationship, weber2020body}.
}

Immersive visualizations commonly employ virtual objects as referents to convey scale perception for abstract data, with scale perception influenced by size perception.
Despite their prevalence, how to set up the design of referents has received limited attention in the literature.
This gap highlights the need for research specifically dedicated to exploring the impacts of referent design on scale perception of the data in immersive visualization.
This study addresses the gap by investigating the impacts of referent layout and referent size on the perception of different data scales.

\subsection{Referent Design in Immersive Visualization}
In the realm of immersive visualization, referents, also known as anchors, are pivotal for providing mental benchmarks that enable intuitive comparisons across data scales~\cite{lee2020data, yang2023understanding}.
\re{
    This concept differs from ``referent'' in situated visualization, a physical object that the data refers to~\cite{willett2016embedded,lee2023design,luo2023pearl}.
}
Chevalier et al.~\cite{chevalier2013using} offered a thorough examination of the referent design space, emphasizing the necessity of considering both the intrinsic properties of the referent object and its relation with the visualization.
In the scope of this research, we consider the intrinsic property of referent size.
\re{
    Research indicates that scale perception is influenced by the dimensions of an object,  with observations showing that enlarging an avatar's hand can cause an underestimation of scale~\cite{ogawa2018object}.
    However, the impact on scale perception when employing external referents within an environmental context remains uncertain.
    Furthermore, there is scant literature on the relationship between referents and visualization within the context of scale perception.
    Therefore, this study draws on concepts from the situated visualization design space, emphasizing the significance of layout as a fundamental link between the referent and the visualization~\cite{willett2016embedded,lee2023design,luo2023pearl}.}
Two prevalent layout designs are considered:  \emph{side-by-side}, where the referent is positioned alongside the visualization but remains distinct, and \emph{in-situ}, wherein the visualization is closely integrated with the referent~\cite{javed2012exploring,chen2017exploring, laviola2023situ}.
While research has evaluated the efficacy of these layouts in specific visual analysis tasks~\cite{wen2022effects,mota2022comparison}, there is limited evidence regarding their impact on data scale perception in immersive visualizations.
\section{Experiment}
\label{sec:study_design}
The study aims to answer the following research question:
\emph{How significantly do design parameters of the referent affect a) the accuracy of scale perception, b) the efficiency of scale perception, and c) the confidence}.
We conduct a within-subjects experiment to address the questions.

\subsection{Study Design}
\subsubsection{Conditions}
The study employs a within-subjects design, focusing on \re{design parameters} including referent layout (\emph{in-situ} vs. \emph{side-by-side}), referent size (\emph{small} vs. \emph{medium} vs. \emph{large}), across small to large data scales.

\begin{figure*}[htb]
      \centering
      \includegraphics[width=0.95\linewidth]{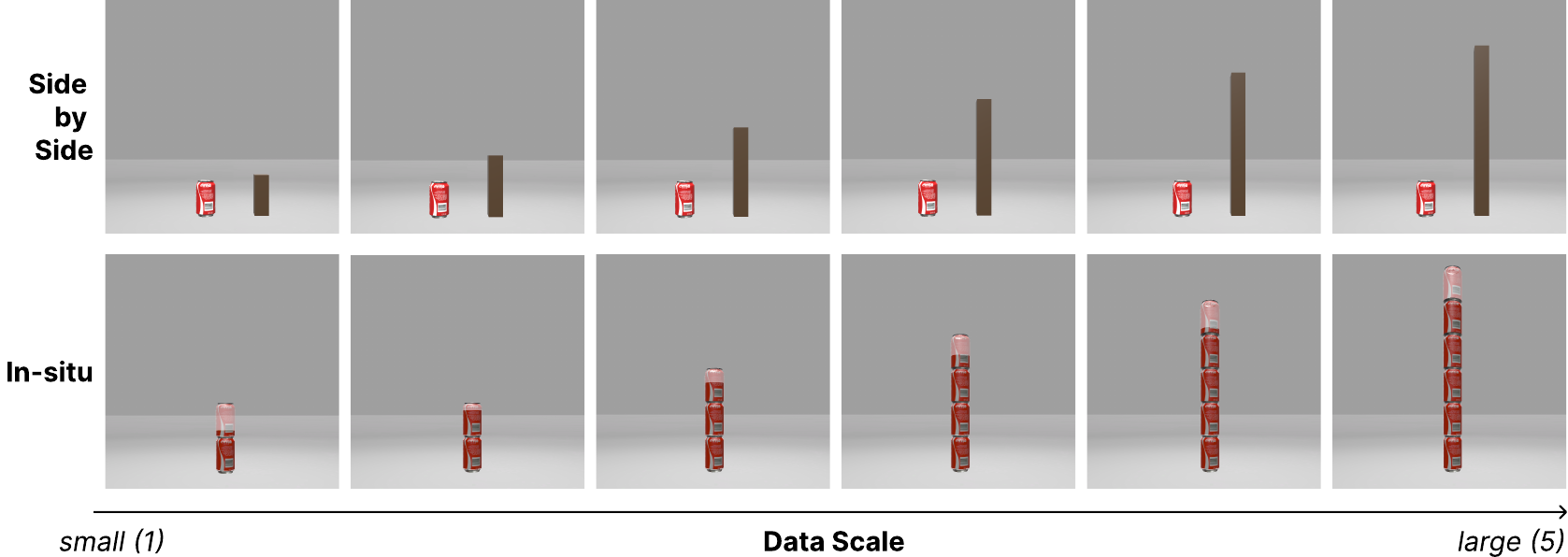}
      \caption{Example stimuli of the experiment, with two referent layouts (side-by-side vs. in-situ) across data scales.}
      \Description{
            The figure shows example stimuli of the experiment, with two referent layouts (side-by-side vs. in-situ) across data scales from 1 to 5.
      }
      \label{fig:scale}
\end{figure*}

\noindent \textbf{Referent Layout:}
We use two layout settings with different integration levels between the referent and visualization, as shown in Figure~\ref{fig:design_parameter} (left).
This followed the classification of layout between visualization and referents proposed by previous studies~\cite{willett2016embedded, laviola2023situ,lee2023design,luo2023pearl}
\re{, as well as real-world practices in immersive storytelling~\cite{yang2023understanding}.
}
Specifically, the setting of the referent layout is as follows:
\begin{itemize}
      \item \textbf{Side-by-side:}
            The referent is positioned adjacent to the visualization, with the visualization represented by a cube on the right and the referent, a realistic object, placed on the left.
            The visualization and referent are clearly separated and placed at the same distance with no obstruction to ensure that the participant can see both the visualization and the referent at the same time.
      \item \textbf{In-situ:}
            The referent functions as a container with the visualization placed inside, serving as the unit of measurement for the visualization.
            The referent's transparency is determined in a way that ensures the visibility of the visualization while distinctly separating it from the environment.
            As height is utilized to encode the data, multiple referents will be stacked to ensure the visualization's height is contained.
            \re{
                  The number of stacked referents matches the data scale, consistent with real-world immersive data visualization storytelling practices, where items like stacked coffins visually signify data magnitudes~\cite{yang2023understanding}.
            }
\end{itemize}

\noindent \textbf{Referent size:}
When selecting referents, a primary consideration is the size, as studies have shown that it is easier for participants to understand familiar sizes than extreme sizes~\cite{chevalier2013using}.
It is important to know whether the referent size will affect the human's ability to estimate the data scale of a visualization.
Therefore, we choose three distinct referent sizes, each benchmarked against the human body size:

\begin{itemize}
      \item \textbf{Small:}
            A soda can, chosen for its familiarity and size approximation to a human hand, stands at a height of 0.12 meters.
      \item \textbf{Medium:}
            A gasoline tank, was selected for its intermediate size of 0.85 meters in height, which closely aligns with a human body size.
      \item \textbf{Large:}
            A shipping container represents a large-size referent with a height of 2.4 meters.
\end{itemize}

\begin{figure}[htb]
      \centering
      \includegraphics[width=0.985\linewidth]{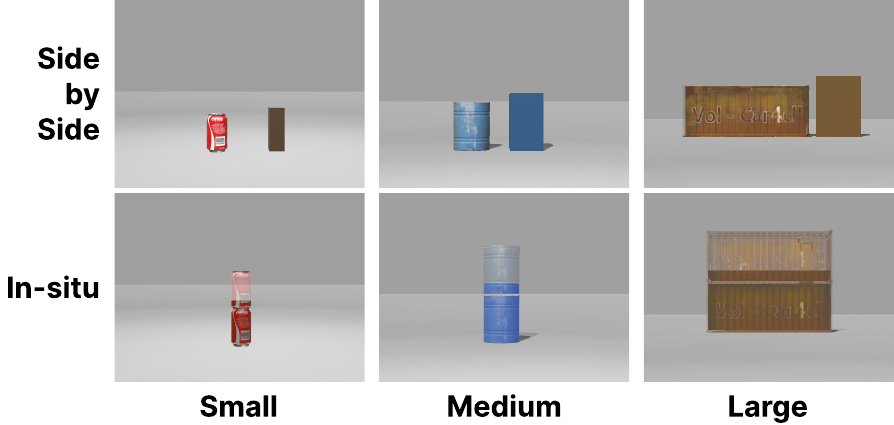}
      \caption{\re{Example stimuli of the experiment, with two referent layouts (side-by-side vs. in-situ) across three referent sizes (small, medium, large).}}
      \Description{
            The figure shows example stimuli of the experiment under three referent sizes (small, medium, large) and two referent layouts (side-by-side vs. in-situ). Details are provided in the text.}
      \label{fig:stimuli}
\end{figure}

To explore the impact of design parameters on performance across varying complexity levels, we conduct experiments with all permutations of referent layout and scale across different data scales, as shown in Figure~\ref{fig:scale}.
The data scale is defined as the ratio between the visualization and the referent.
A smaller ratio is relatively easy for comparisons as the visualization size is close to the referent size, while a larger ratio increases complexity.
Note that when the data scale increases, the in-situ layout will use multiple stacked referents, while the side-by-side layout will use a single referent.
\re{In our study, we adopted a 1 to 5 continuous data scale with random numbers for precise participant feedback capture aimed at investigating scale perception nuances.}
The range was informed by a preliminary study, which indicated that data scales below 1 yield representations too small for accurate control, and scales above 5 create visualizations too large for the head-mounted display's (HMD) field of view.

\subsubsection{Stimuli}
\re{
      The stimuli consist of a visualization and a referent, with the visualization representing the data scale and the referent serving as a benchmark for scale perception, as shown in Figure~\ref{fig:stimuli}.
      To minimize cognitive errors associated with environmental textures, the scene was rendered in a simplistic gray-and-white palette, complemented by high-fidelity rendering techniques to improve scale perception accuracy in VR, as supported by research~\cite{ogawa2019virtual, wijayanto2023comparing}.
      Previous study suggests that the perception of object size is independent of object distance when using familiar size cues, where the clarity and recognition of the referent by participants is more crucial~\cite{haber2001independence}.
      Accordingly, stimuli were positioned at distances of 1, 6, and 10 meters, following pilot feedback, to ensure visibility and recognition without exiting the field of view of the Head-Mounted Display (HMD).
      Similarly, to improve visibility, a cola can is suspended in mid-air instead of being placed on the ground.
}

\subsubsection{Measures}
The experiment gauges participant performance and subjective responses across several dimensions:

\begin{itemize}
      \item \textbf{Scale ratio:}
            Consistent with previous works~\cite{ogawa2019virtual, arora2021thinking}, the scale ratio is employed as a metric for accuracy.
            This ratio is calculated as the proportion between the perceived scale and the actual scale of the visualization cube.
            A scale ratio of 1 indicates perfect accuracy, denoting an exact match between the perceived and the actual data scale.
      \item \textbf{Time consumption:}
            Participant completion times for the size-matching task are measured from the onset of the recall stage to the submission of their response.
      \item \textbf{Confidence:} Participants' self-reported confidence in their ability to complete the size matching task is assessed using a 5-point Likert scale ranging from 1 (poor) to 5 (excellent).
\end{itemize}

\subsubsection{Hypotheses}
Based on previous research, we formulate the following hypotheses:

\noindent \textbf{Accuracy (H1):} Layout and referent size influence the accuracy of scale perception.

\begin{enumerate}
      \item[H1.1] In-situ layout will improve accuracy in scale perception compared to side-by-side layouts.
      \item[H1.2] A large referent size will lead to lower accuracy in scale perception compared to a small referent size.
      \item[H1.3] \re{The effect of referent size on accuracy is moderated by layout type.
                  Specifically, for tasks involving larger referents, an in-situ layout is expected to result in higher accuracy than a side-by-side layout.
                  Conversely, this advantage of in-situ layouts diminishes or reverses with smaller referents.}
      \item[H1.4] An interaction effect is anticipated between referent layout and data scale, with in-situ layout exhibiting more resilient accuracy when data scale increases.
\end{enumerate}

\noindent \textbf{Efficiency (H2):} Layout and referent size influence the efficiency of scale perception.

\begin{enumerate}
      \item[H2.1] In-situ layouts facilitate quicker task completion compared to side-by-side layouts.
      \item[H2.2] Smaller referent sizes lead to faster task completion compared to larger referent sizes.
      \item[H2.3] \re{Similar to H1.3, the effect of referent size on task completion time is moderated by layout type.}
      \item[H2.4] An interaction effect is anticipated between referent layout and data scale, with in-situ layout exhibiting more resilient efficiency affected by data scale.
\end{enumerate}

\noindent \textbf{Confidence (H3)}: In-situ layouts can increase confidence compared to side-by-side layouts.

\re{
      Hypotheses H1.1, H2.1, and H3 propose that in-situ layouts, by providing rich contextual information, facilitate easier comparisons for participants.
      H1.2 and H2.2 suggest that small referent lead to higher accuracy and efficiency due to familiarity.
      H1.3 and H2.3 contend that larger referents pose more challenges in side-by-side layouts because of increased cognitive load, as they lack integration with the visualization.
      Finally, H1.4 and H2.4 argue that, although larger data scales increase cognitive demands, in-situ layouts ease this by adding context.
}

\subsection{Participants}
In the study at hand, 28 individuals were recruited from a university campus.
The demographic composition included 19 males and 9 females with an age distribution between 21 and 30 years (M = 23.6, SD = 2.0).
With respect to VR proficiency, the majority (n=21) reported minimal experience, having engaged with VR for less than one year.
A minority had more extended interactions, with 4 participants in the 1-2 year category, a single participant with 3-4 years of VR experience, and 2 participants exceeding 4 years.
Pertinent to the focus of the study, VR experience specific to visualization tasks varied: 6 participants had no prior exposure, 19 had a cursory acquaintance having tried it a few times, 2 reported frequent usage, and only 1 participant was deemed an expert.

\subsection{Apparatus}
The experimental setup featured a Windows 11 PC, Meta Quest 2 head-mounted display with a 1920 x 1832 per-eye resolution and ~100 degrees FOV, and Oculus hand-held controllers using a Unity 3D-developed VR program.
\re{During the experiments, participants were seated with head movements limited to a 30-degree horizontal range, sufficient for viewing stimuli while minimizing locomotion's impact on scale perception.
      The interpupillary distance (IPD) was defaulted to the average value of 63mm, while participants were permitted to adjust it for a clearer view.
}
Operational data were continuously logged for analysis.

\begin{figure*}[htb]
      \centering
      \includegraphics[width=0.95\linewidth]{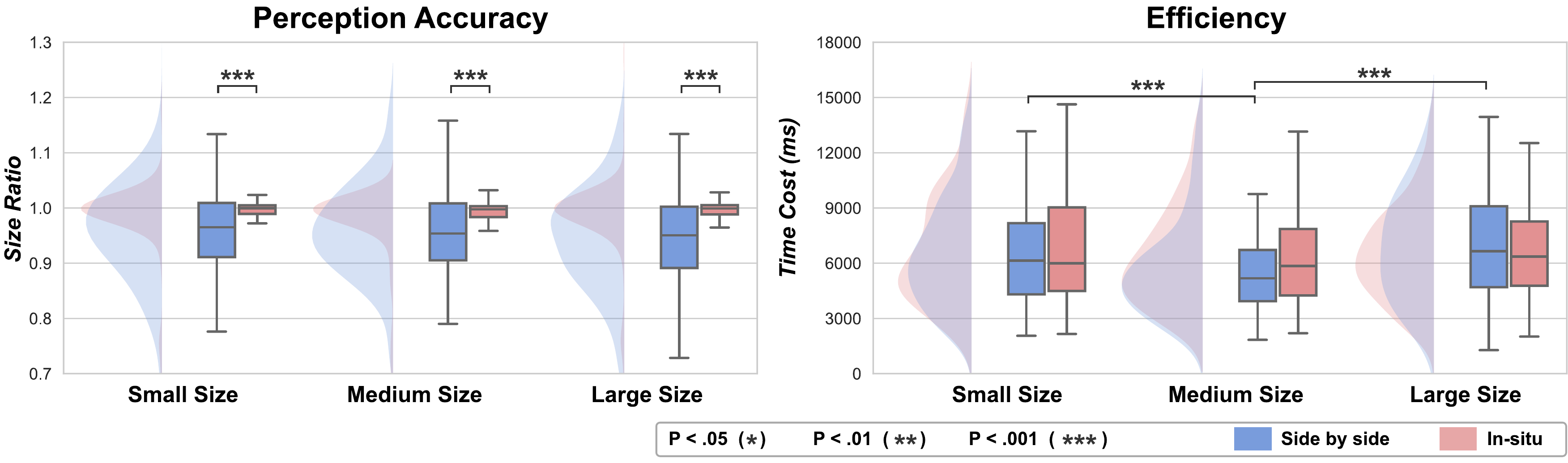}
      \caption{Boxplot of scale perception accuracy and efficiency across different referent layouts and referent sizes.}
      \Description{This figure illustrates the boxplot of scale perception accuracy and efficiency across different referent layouts and referent sizes. Details are provided in the text.}
      \label{fig:result-significants}
\end{figure*}

\subsection{Procedure}
The experiment took place in a quiet room equipped with a desk and chair.
Participants first signed a consent form and provided demographic information before donning the VR headset and controllers.
A brief training session introduced them to the VR environment, controller use, and the two referent layouts, including a practice size-matching task.
Once participants felt prepared, the experiment commenced, with the trial flow structured into sequential steps:
\begin{itemize}
      \item \textbf{Observation Phase:}
            Participants were presented with the visualization cube with referent for a duration of 3 seconds to observe and mentally estimate its height.
            \re{
                  This duration was chosen based on previous research, which suggests that the object size can be recognized in a short time~\cite{isik2014dynamics}, also validated in our pilot study.
            }
      \item \textbf{Interference Phase:}
            Immediately after observation, both the cube and referent were hidden, and visual masking was applied for 2 seconds to eliminate afterimages.
      \item \textbf{Recall and Matching Phase:}
            Participants were then presented with an adjustable 'answer cube' to resize using the VR controllers, aiming to match the observed cube's height.
            They had up to 20 seconds for adjustments, with their final action recorded as the response upon pressing the controller's trigger button.
            \re{This setup aligns with the protocols established in prior research, enabling the accurate collection of data on perceived size \cite{ogawa2019virtual, arora2021thinking}.}

\end{itemize}
\re{Each participant will go through all the conditions, with the order being randomized to mitigate potential learning effects.}
Between trials, participants were allowed rest periods, proceeding to the next trial at their discretion by clicking a button.
Upon completing all trials for a given referent, they completed a post-trial questionnaire assessing their confidence in task completion.
Avoid participants using the initial position as a reference, the initial height of the answer cube was set to a random value calculated based on the middle point of the visualization cube's height $m$ and the referent's height $R_h$, with a random perturbation $\epsilon$ of 20\%:
$$
      V_h=(m + \epsilon)R_h, where \ m =(1+5)/2=3, and \ \epsilon \in [-0.1, 0.1].
$$

\re{The study took approximately 30 minutes to complete, and participants were compensated with 50 RMB for their participation.}

\section{Results}
\label{sec:results}

\re{Each participant completed 4 trials for each of the 3 referent sizes and 2 layouts, resulting in a total of 672 trials involving 28 participants.}
Trials were subject to exclusion based on two criteria: size ratio and completion time, specifically those deviating by more than two standard deviations from the mean.
This exclusion was aimed at eliminating trials potentially compromised by inadvertent button clicks or non-completion of tasks.
This led to the exclusion of 24 trials, accounting for 3.57\% of the total.
For the remaining trials, we first use the Kolmogorov-Smirnov test to check the normality of the data.
Then due to the non-normality of the data, we performed the Wilxocon signed-rank test and the Fridman test to assess task accuracy and efficiency, followed by Wilxocon post-hoc tests for significant pairwise comparisons of data.

\subsection{Accuracy}
We conducted statistical analyses on the size ratio to determine the effects of reference layout, referent size, and data scale on size perception accuracy.
The analysis utilized the average size ratio across four trials for each of the six condition combinations per participant.
The in-situ layout yielded significantly higher accuracy than the side-by-side layout (\( p < .001 \)).
As shown in Figure~\ref{fig:result-significants}, when analyzing the data grouped by the referent size, all three referent sizes demonstrated significantly higher accuracy with the in-situ layout compared to the side-by-side layout (\( p < .001 \)), leading to the acceptance of hypothesis H1.1.
The overall size ratio of the in-situ layout is 1.52\% higher than the side-by-side condition (M=0.985, SD=0.004 vs. M=0.942, SD=0.009).
This suggests that the in-situ layout is more conducive to accurate size estimation, regardless of the referent size.
\re{
    No significant differences were observed for referent sizes or between layout and referent size, leading to the rejection of hypotheses H1.2 and H1.3.
}

\begin{figure}[htb]
    \centering
    \includegraphics[width=0.985\linewidth]{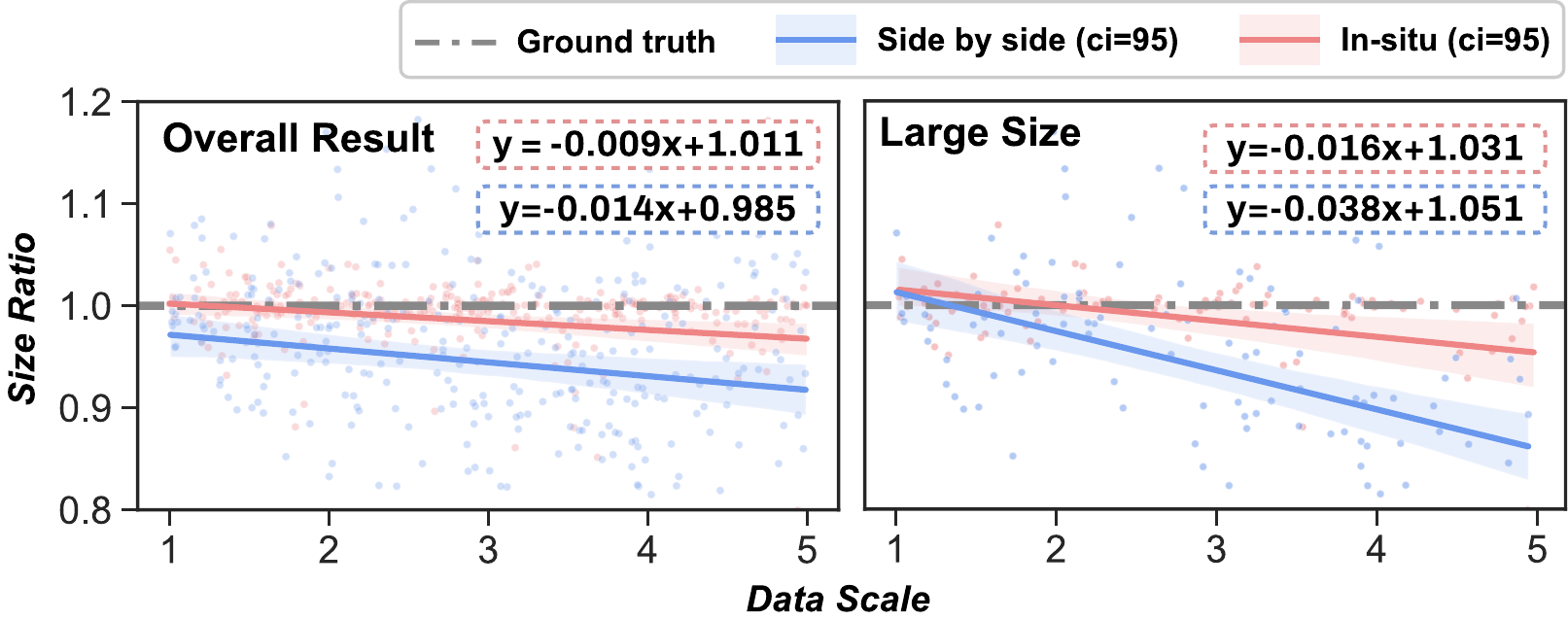}
    \caption{Scatter plot showing linear regression profiles of the different layouts in the relationship between data scale (predictor) on size ratio (predicted). The Left shows the overall regression profiles, while the right shows the regression profiles for large-size referents.}
    \Description{
        The figure shows two scatter plots of linear regression profiles of the different layouts in the relationship between data scale (predictor) on size ratio (predicted). The regression equations are provided in the text.
    }
    \label{fig:accuracy}
\end{figure}

In order to conduct multiple regression analysis and evaluate how layout, data scale, and referent size affect the size ratio, a multiple regression was calculated to predict the size ratio based on the independent variables.
The regression model on size ratio was found to be significant, F(3, 644) = 14.756, ($p < 0.001$), with $R^2 = 0.06$.
The size ratio was significantly predicted by layout ($p < 0.001$), data scale ($p < 0.001$), and no significant effect of referent size.
The relationship between the size ratio and the independent variables was determined to be:
$Size \ Ratio = 102\% - 1.1\% \times Data \  Scale - 4\% \times Layout$.
This indicates that for every 1 unit increase in data scale, the size ratio decreased by 1.1\%, and a 4\% difference in size ratio was observed on average between two subsequent layout conditions.
For each layout, linear regression analysis was then conducted, as shown in Figure~\ref{fig:accuracy}.
For the in-situ layout, the equation obtained was $Size Ratio = 101.1\% -0.9\% \times Data \ Scale$, while for the side-by-side layout, it was $Size Ratio = 98.5\% - 1.4\% \times Data \ Scale$, shown in Figure~\ref{fig:accuracy}.
The findings revealed an underestimation of size in VR, consistent with previous studies~\cite{thompson2004does, kenyon2007size, kelly2017perceived}.
Furthermore, when employing a large-size referent, the difference between the two layouts became more pronounced.
The in-situ layout exhibited a 1.6\% distortion for each unit increase in data scale, whereas the side-by-side layout displayed a 3.8\% distortion for each unit increase in data scale, accepting hypothesis H1.4.

\subsection{Efficiency}
We statistically tested the completion time to examine whether the task completion time was influenced by the referent layout, referent size, and data scale.
Figure~\ref{fig:result-significants} illustrates the results.
No significant difference was found between the in-situ layout (M=6630.8, SD=2184.2) and the side-by-side layout (M=6470.3, SD=2296.8), leading to the rejection of hypothesis H2.1.
For the referent size, the completion time of the medium size (M=6054.2, SD=2084.9) is significantly shorter than the small size (M=6780.3, SD=2245.8) (\( p < .001 \)), and large size (M=6826.0, SD=2323.5) (\( p < .001 \)), leading to the reject of hypothesis H2.2.
\re{
    No interaction effect was found between layout and referent size, leading to the rejection of hypothesis H2.3.
}

The regression model on completion time was found to be significant, F(3, 646) = 3.413, ($p < 0.005$), with $R^2 = 0.011$.
Through multiple regression analysis, we found that the completion time was significantly predicted by the data scale ($p < 0.005$), and no significant effect of layout and referent size.
The relationship between completion time and the independent variables was determined to be:
$Completion \ Time = 5492.9ms + 295.5ms \times Data \ Scale$, which indicates that for every 1 unit increase in data scale, the completion time increased by 295.5ms.
For each layout, linear regression analysis was then conducted, and the effects are both significant ($p < 0.05$).
For the in-situ layout, the equation obtained was $Completion \ Time = 5717.5ms + 274.0ms \times Data \ Scale$, while for the side-by-side layout, it was $Completion \ Time = 5533.8ms + 296.6ms \times Data \ Scale$, leading to the reject of hypothesis H2.4.
Despite the in-situ layout's larger constant term compared to the side-by-side layout, its smaller slope indicates that completion times in the in-situ layout are less impacted by data scale changes.

\subsection{User Feedback}
Task completion confidence was notably higher with the in-situ layout (M=4.13, SD=0.67) compared to the side-by-side layout (M=3.06, SD=0.72), accepting H3.1 with 88\% of participants responding positively for the in-situ layout against 23.8\% for the side-by-side layout.

\begin{figure}[htb]
    \centering
    \includegraphics[width=0.985\linewidth]{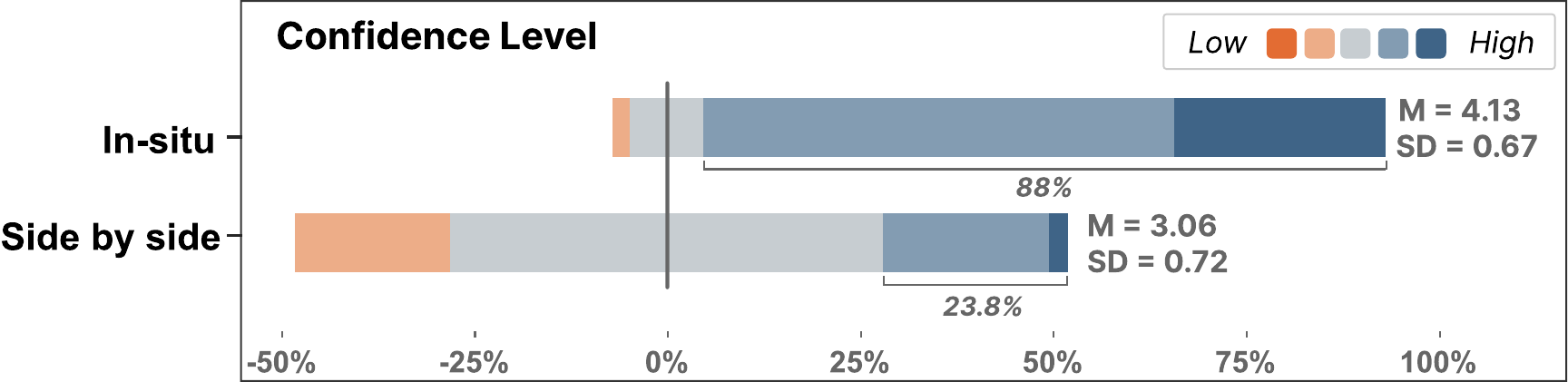}
    \caption{Users' self-reported confidence on the size-matching task completion for two layouts. The stacked bar chart shows the percentage of participants for each confidence level.}
    \Description{
        The figure shows a stacked bar chart of users' self-reported confidence in the size-matching task completion for two layouts. The in-situ layout has a higher percentage of participants with higher confidence levels compared to the side-by-side layout.}
    \label{fig:feedback}
\end{figure}

\section{Discussion}
\textbf{Findings}.
The regression profiles for layout impact on accuracy across data scales revealed significant advantages of the in-situ layout, which is attributed to two main factors.
First, it integrates referents and visualizations, reducing the need for mental computation. 
This is supported by its regression constant being closer to 1, indicating a higher base accuracy level.
Second, the in-situ layout demonstrates a smaller regression slope than the side-by-side layout, indicating less sensitivity to data scale changes and greater resilience to size distortions.
This resilience may stem from enhanced depth cues, such as texture gradients, which are more effectively preserved in the in-situ layout. 
This effect is more obvious when using a large referent, since for the same data scale, a large referent may contain more depth cues, thus the lack of depth cues is more significant.

Another interesting finding is that using medium-sized referents is more efficient compared to small and large referents. 
This could be because the human visual system is particularly attuned to objects of a size similar to that of the human scale, facilitating quicker perception. 
Smaller referents lead to increased adjustment and confirmation time, despite accurate size perception, due to the heightened visual acuity needed for detail. 
\re{Larger referents may also overwhelm the visual field, thus extending the time required for thorough perception and integration.}

\noindent
\textbf{Design Suggestions}.
Based on our findings, we advocate for the use of in-situ layouts for visualizations across varying referent sizes, as they have demonstrated superior performance in enhancing accuracy and confidence compared to side-by-side layouts. 
The resilience of in-situ layouts to changes in data scale makes them particularly advantageous for large-scale visualizations. 
Furthermore, we recommend employing referents that approximate the human scale, especially in contexts where efficiency is paramount. 
This recommendation is grounded in the observation that objects of a size similar to the human body are perceived more efficiently, owing to greater familiarity and perceptual efficiency with such scales.
Implementing these strategies can notably boost the effectiveness and efficiency of immersive visualization systems. 
\re{This approach is particularly beneficial in immersive data storytelling, where the rapid and accurate conveyance of abstract data is essential for effective comprehension and interpretation.}

\noindent
\textbf{Limitations and Future Work}.
Our study introduces certain limitations that naturally open avenues for future research endeavors. 
Specifically, our examination of referent design was confined to aspects of layout and size, neglecting the exploration of more intricate strategies. 
These overlooked strategies include the deployment of semantically continuous referents across various scales and the utilization of multiple smaller referents to represent larger quantities, as discussed in prior work~\cite{chevalier2013using}. 
Additionally, our analysis was solely concentrated on the use of height for encoding information within visualizations, leaving the potential of alternative encoding techniques such as area and volume unexamined.
\re{
    Our study primarily focuses on the visual design of referent in VR and does not delve into how interactivity, like walking around, might influence design choices~\cite{creem2023perceiving}. 
    We limit participants to minimal movements, with our analysis concentrating solely on the effects of dynamic viewing angles. 
    The potential impact of broader interactive behaviors, such as locomotion, remains unexplored in this context.
}
Future investigations could extend into these areas, enriching our understanding of effective referent design in immersive visualizations.
\section{Conclusion}
\label{sec:conclusion}
This study examines the impact of referent design on scale perception in immersive visualizations.
Through comparative analysis of layout and referent size, findings indicate that in-situ layouts notably improve accuracy and confidence in scale perception, particularly at larger data scales, with medium-sized referents facilitating quicker task completion. 
These results offer essential guidance for referent design in immersive environments.

\bibliographystyle{ACM-Reference-Format}
\bibliography{sample-base}



\end{document}